\documentclass[]{interact}

\usepackage{epstopdf}% To incorporate .eps illustrations using PDFLaTeX, etc.
\usepackage[caption=false]{subfig}% Support for small, `sub' figures and tables

\usepackage[numbers,sort&compress,merge]{natbib}% Citation support using natbib.sty
\bibpunct[, ]{[}{]}{,}{n}{,}{,}% Citation support using natbib.sty
\theoremstyle{plain}% Theorem-like structures provided by amsthm.sty

\theoremstyle{definition}

\theoremstyle{remark}

\usepackage[version=4]{mhchem}
\usepackage{color}
\usepackage{lineno}

\usepackage{xspace}%adds a space when needed
\newcommand{\hcop}{HC$_3$O$^+$\xspace}
\newcommand{\hcsp}{HC$_3$S$^+$\xspace}
\newcommand{\wn}{cm$^{-1}$\xspace}
\usepackage{float}

\usepackage{xcolor}
\usepackage{textcomp}

\begin{document}

%\linenumbers

\articletype{Article} % Specify the article type or omit as appropriate

\title{Descendant of the X-ogen carrier and a ``mass of 69'':\\ Infrared action spectroscopic detection of \hcop and \hcsp \thanks{{This article has been accepted for publication in ``Molecular Physics'', published by Taylor and Francis}}} %\mam{I think HCO+ should appear in the title together with X-ogen, because Mol Phys reader won't have any idea of what it means. NO!}}

\author{
\name{
Sven Thorwirth,\textsuperscript{a}\thanks{CONTACT S. Thorwirth. Email: sthorwirth@ph1.uni-koeln.de} 
Michael~E. Harding,\textsuperscript{b}\thanks{CONTACT M.~E. Harding. Email: michael.harding@kit.edu}
Oskar Asvany,\textsuperscript{a}
Sandra Br{\"u}nken,\textsuperscript{c}
Pavol Jusko,\textsuperscript{a*}\thanks{*Present Address P. Jusko. Max Planck Institute for Extraterrestrial Physics, Gie{\ss}enbachstraße 1, 85748 Garching, Germany}
Kin Long Kelvin Lee,\textsuperscript{d}
Thomas Salomon,\textsuperscript{a}
Michael C. McCarthy,\textsuperscript{d}
and Stephan Schlemmer\textsuperscript{a}
}
\affil{
\textsuperscript{a} I. Physikalisches Institut, Universit\"at zu K\"oln, Z\"ulpicher Str. 77, 50937 K\"oln, Germany\\
\textsuperscript{b} Institut f\"{u}r Physikalische Chemie, Abteilung f\"ur Theoretische Chemie, Karlsruher Institut f\"{u}r Technologie (KIT), Kaiserstraße 12, 76131 Karlsruhe, Germany\\
\textsuperscript{c} Radboud University, Institute for Molecules and Materials, FELIX Laboratory, Toernooiveld 7, 6525ED, The Netherlands\\
%\textsuperscript{d} Max Planck Institute for Extraterrestrial Physics, Gie{\ss}enbachstraße 1, 85748 Garching, Germany;
\textsuperscript{d} Center for Astrophysics $|$ Harvard \& Smithsonian, 60 Garden St., Cambridge, MA 02138, U.S.A.
}}

\maketitle

\begin{abstract}
The carbon chain ions \ce{HC3O+} and \ce{HC3S+} -- longer variants of the 
famous ``X-ogen'' line carrier \ce{HCO+} - have been observed for the first time
using two cryogenic 22-pole ion trap apparatus (FELion, Coltrap) and two different light sources: the Free Electron Laser for Infrared eXperiments (FELIX), which was operated between 500 and 2500\,\wn, and an optical parametric oscillator operating near 3200\,\wn; signals from both experiments were detected by infrared predissociation action spectroscopy.
The majority of vibrational fundamentals were observed for both ions and their
vibrational wavenumbers compare very favorably
with results from high-level anharmonic force field calculations performed here at the coupled-cluster level of theory. As the action spectroscopic scheme  probes
the Ne-tagged weakly bound variants, Ne$-$\hcop and Ne$-$\hcsp,  corresponding calculations of these systems were also performed. Differences in the structures and molecular force fields between the bare ions and their Ne-tagged complexes are found to be very small.
%All infrared active fundamental vibrational modes in the range 500 to 2500~cm$^{-1}$ have been 
%identified, supported by high-level quantum-chemical computations employing highly-accurate coupled-cluster computations 
%together with large basis sets. 
\end{abstract}


\section{Introduction}

Acylium ions, R$-$C$\equiv$O$^+$, play a vital role as reaction intermediates in preparatory organic synthesis, most notably as electrophiles in coupling reactions such as the Friedel-Crafts acylation \cite[see, e.g., Ref.][]{carey_sundberg}. %Also, acylium 
These ions and their sulfur analogs, R$-$C$\equiv$S$^+$, have also been invoked in astrochemical reaction networks 
%calculations
to account for the production of (carbon-rich) cumulenic chains via very fast dissociative recombination (DR) with free electrons %, e.g.\mam{,}
\cite[e.g., Ref.][]{mcelroy_AA_550_A36_2013}
%\mam{:} \mamcomm{Isn't DR always with free electrons? Otherwise shouldn't it be "dissociative recombination (DR) with free electrons reactions"}%,

%\begin{equation}\label{eq:DR}
%    \rm HC_3O/S^+ + e^- \to C_3O/S + H.
%\end{equation}

%\mam{proposition:}
%\mam{
\begin{equation}
    \ce{HC3}X^+ + \mathrm{e}^- \to \ce{C_3}X + \mathrm{H},  X = \mathrm{O, S}.
    \label{eq:DR}
\end{equation}

\noindent
%Indeed, 
The astronomical identification of \ce{C3O} and \ce{C3S}, both products of reaction \eqref{eq:DR}, nearly 35 years ago \cite{matthews_nature_310_125_1984,yamamoto_ApJ_317_L119_1987} points to the importance of the parent ions in chemical networks of dark molecular clouds.

The prototypical acylium --- formylium ion, \ce{HCO+} --- was the first polyatomic ion detected in space, although its identity was originally a mystery when a strong line at 89.2\,GHz was observed towards several astronomical objects; for lack of a better name, the line was dubbed ``X-ogen'' \cite{buhl_nature_228_267_1970} (``extraterrestrial origin"). However, it was not until the
pure rotational spectrum of this ion was measured by Woods and co-workers a few years later \cite{woods_PRL_35_1269_1975} that the carrier of the astronomical line was established with certainty, and in doing so the bold prediction of Klemperer was confirmed \cite{klemperer_Nature_227_1230}. 
Similarly, \ce{HCS+} was first detected in space \cite{thaddeus_ApJ_246_L41_1981} almost simultaneously with a report of its high-resolution laboratory spectrum  \cite{gudeman_ApJ_246_L47_1981}.
Since then, \ce{HCO+} and \ce{HCS+} have been the subjects of a number of high-resolution spectroscopic investigations %given 
 %on both \ce{HCO+} and \ce{HCS+} 
\cite[][]{lattanzi_ApJ_662_771_2007,tinti_ApJ_669_L113_2007,cazzoli_ApJSS_203_11_2012,siller_JPCA_117_10034_2013,rosenbaum_JMSt_133_365_1989,tang_ApJ_451_L93_1995,margules_PCCP_5_2770_2003}.

While many mass spectrometric studies involving acylium and thioacylium ions have been reported \cite[see, e.g., Refs. ][]{mass_spectrometry_acylium,caserio_JACS_23_6896_1983,rahman_OrgMassSpectrom_23_517_1988,liu_JPCA_101_4019_1997,liu_IJMSIP_155_79_1996,pepp_JPCA_104_5817_2000,derbali_PCCP_21_14053_2019}, gas-phase spectroscopic investigations beyond the simplest members of each family, \ce{HCO+} and \ce{HCS+}, %HCO/S$^+$ 
%with $X\neq \rm H$ 
are scarce. 
Thioacylium species appear to have only been studied using mass spectrometry while gas-phase spectroscopy of acylium ions is limited to only a few  species. The acetyl cation, \ce{CH3CO+}, for example, has been studied in a free-jet expansion source by infrared (IR) photodissociation  of its  van der Waals complex \ce{CH3CO+}$-$Ar \cite{mosley_JCP_141_024306_2014}. An electronic spectrum of \ce{HC7O+} in the gas-phase and trapped in an inert gas matrix has also been  obtained \cite{chakraborty_MolPhys_114_2794_2016}. Very recently, IR photodissociation spectra of \ce{HC_nO+}$-$CO complexes ($n=$ 5$-$12) have been reported in the 1600 to 3500 \ce{cm^{-1}} region \cite{jin_JCP_146_214301_2017,li_CJCP_32_77_2019}. %\cite{liu_IJMSIP_155_79_1996,liu_JPCA_101_4019_1997}.

This paper reports the first spectroscopic observations of the \hcop and \hcsp ions. Experimental measurements were performed at IR wavelengths by making use of sensitive action spectroscopic techniques in combination with modern ion traps.  
%Spectra were recorded from 500 to 2500\,\wn at the %with 
%Free Electron Laser for Infrared eXperiments (FELIX) facility (Radboud University, Nijmegen), with follow-up studies near 3200\,\wn using an optical parametric oscillator (University of Cologne).
Spectroscopic assignment was based on high-level quantum-chemical calculations performed at the CCSD(T) level of theory \cite{raghavachari_chemphyslett_157_479_1989}, in conjunction with the experimental work.
%It should be noted that \hcop is isoelectronic with cyanoacetylene,
%\ce{HC3N}, a molecule that has been characterized in the infrared very well \cite{Jolly_JMS_242_46_2007}.
%Due to this relationship, \hcop and \ce{HC3N} should exhibit
%similar spectroscopic characteristics. The isoelectronic variant of \ce{HC3S+}, phosphabutadiyne, \ce{HC3P}, has not yet been studied in the %infrared.

%So far, 

%%%%%%%%%%%%%%%%%%%%%%%%%%%%%%%%%%%%%%%%%%%%%%%%%%%%%%%%%%%%%%%%%%%%%%%%%%%%%%%%%%%%%%%%%%%%%%%%%%%%%%%%%%%%%%%%%%%%%%%%%%%%%

\section{Experiment}

Experimental characterization of \hcop\ and \hcsp\ in the wavenumber range from 500 to 2500~\wn\
was performed in the cryogenic 22-pole ion trap
apparatus FELion connected to the Free Electron Laser for Infrared eXperiments, FELIX~\cite{oepts_IPT_36_297_1995}, located at Radboud University (Nijmegen, The Netherlands).
The FELion apparatus has been described in detail recently \cite{jus19}.
Briefly, primary ions are produced using electron impact ionization of suitable precursor compounds and electron energies of a few tens of eV and source pressures in the $10^{-5}$~mbar regime are common. Subsequent reactions in the storage ion source may lead to the formation of secondary ions.
A pulse of ions is then extracted from the source into a first quadrupole mass filter to select the ion mass of interest, $A^+$. 
These ions are then guided into a 22-pole ion trap~\cite{asv10} where
they collide with a bath of rare gas (rg, typically Ne or He) provided as an intense pulse at the beginning of the trapping cycle. 
Collisions with rg atoms inside the trap cool the ions both kinetically and internally. Provided the temperature is sufficiently low, of order a few K, 
van der Waals clusters of ions and the rg will form by three-body collisions. 
Using a second mass filter stage connected to the
exit of the 22-pole ion trap, the ions and their  complexes can be mass filtered and detected using 
a very sensitive Daly type detector after a selected storage time. In the absence of any  chemical reactions in the trap, the
ion distribution exiting the trap includes the bare ion $A^+$, 
with smaller amounts of weakly bound van der Waals clusters of 
the ion with one or more rg atoms, $A^+-$rg$_n$~\cite{bru03,jas13}. 
In the infrared predissociation (IRPD) action spectroscopy scheme  employed here, 
the number of singly tagged  $A^+-$rg cluster ions is constantly monitored 
while the FELIX (FEL-2) IR radiation traversing the ion trap is tuned in wavenumber. 
When a vibrational mode of the cluster and the radiation source are coincident in wavenumber, dissociation of the cluster occurs, resulting in depletion in the $A^+-$rg counts at the detector. Since the weakly bound rare gases He or Ne generally only slightly perturb the structure of the ion, the IRPD spectrum is highly representative of  the nascent ion.
Using FELion in combination with FELIX, the IRPD approach has been used successfully in recent years to infer infrared spectra for many fundamental ions~\cite{bru19,jus19,asv19,kohguchi_JCP_148_144303_2018}.

\begin{figure}[H]
\centering
\includegraphics[width=15cm]{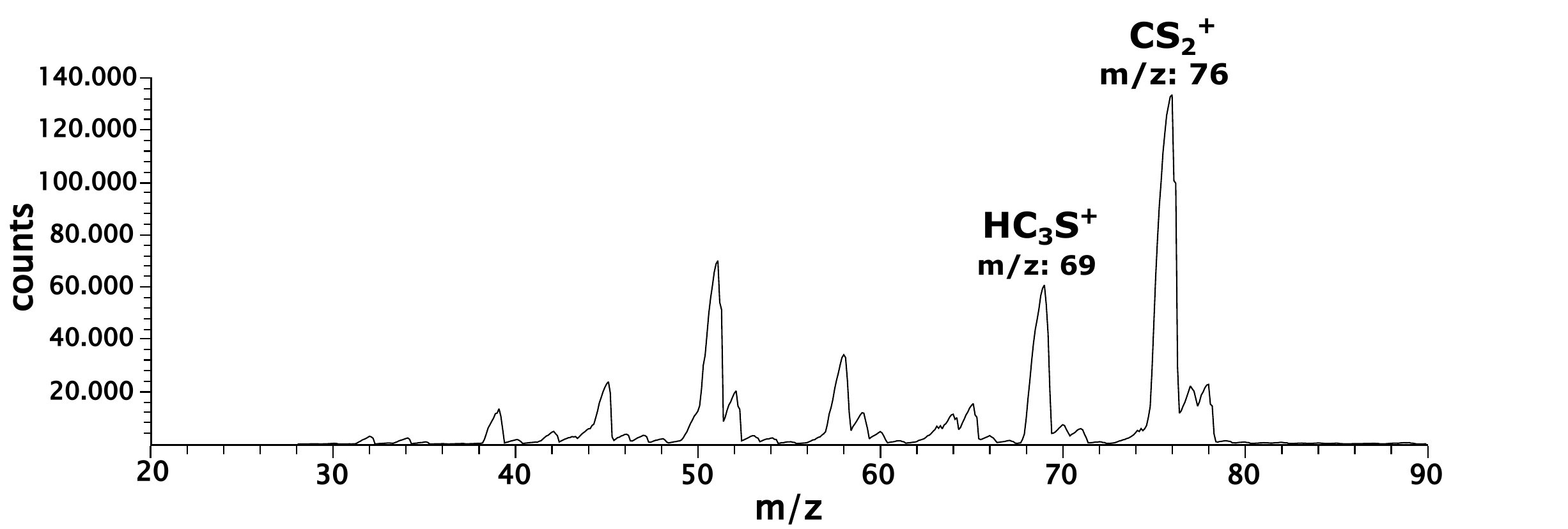}
\includegraphics[width=15cm]{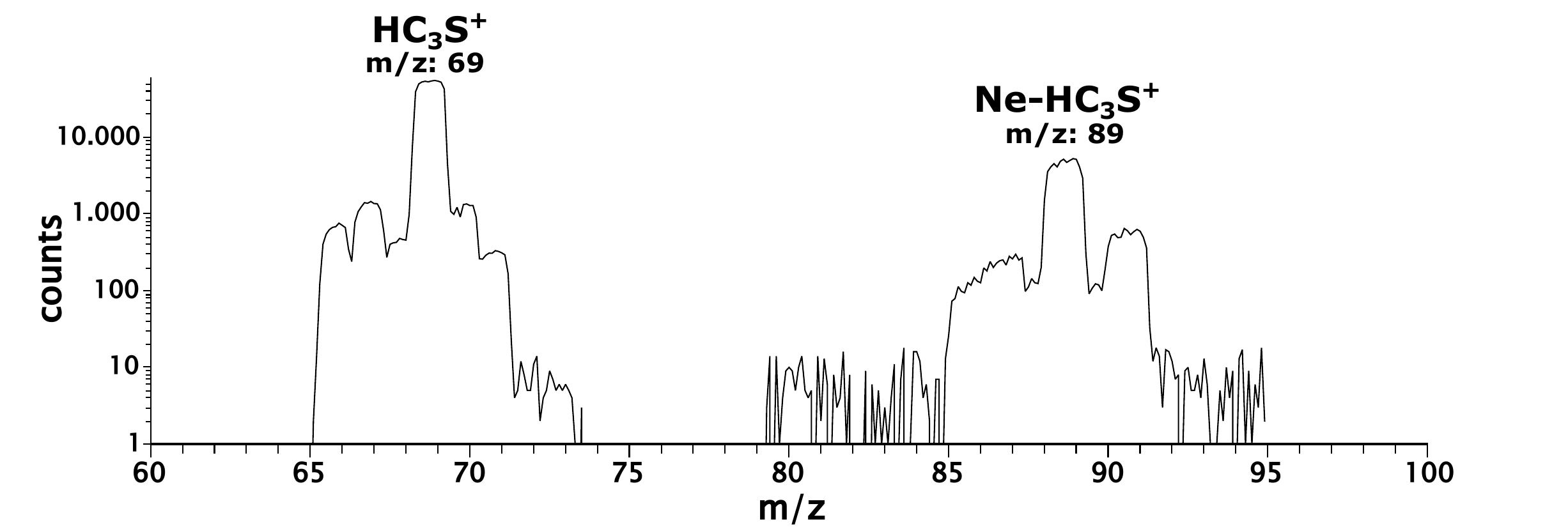}
%%\subfloat[An example of an individual figure sub-caption.]{%
%%%\resizebox*{5cm}{!}{\includegraphics{graph1.eps}}}\hspace{5pt}
%%\subfloat[A slightly shorter sub-caption.]{%
%%%\resizebox*{5cm}{!}{\includegraphics{graph2.eps}}}
\caption{Mass spectra obtained from electron impact ionization of an acetylene/\ce{CS2} mixture. Top: Unbiased scan of source content: $m/z=76$ is from the \ce{CS2+} ion, counts of $m/z=69$ from \ce{HC3S+} are about 50~\% that of \ce{CS2+}. $m/z=51$ is most probably from a \ce{C4H3+} ion. Bottom: Logarithmic plot of trap content extracted from the cold trap (8~K) after storing  mass filtered \ce{HC3S+} ($m/z=69$) for 0.5~s (some neighbouring masses are observed at low counts due to somewhat course mass selection conditions). In addition to the $m/z=69$ peak at about 50000 counts, some 5000 counts of the Ne$-$\ce{HC3S+}
cluster are detected at $m/z=89$. 
} \label{mass_spec_hc3sp}
\end{figure}

The results presented here were obtained during two independent observing campaigns, carried out at the FELIX laboratory in 2016 (\hcop ) and 
2018 (\hcsp ).
Target ions were produced 
through electron impact ionization of simple commercially available 
precursors. 
\hcop\ was produced via dissociative ionization of propargyl alcohol (H$-$C$\equiv$C$-$CH$_2$OH, Sigma-Aldrich) 
and \hcsp\ from ionization and secondary reactions of a mixture of acetylene and \ce{CS2}. Following mass selection
in the first quadrupole stage,
HC$_3$O$^+$ ($m/z=53$) or HC$_3$S$^+$ ($m/z=69$) 
were introduced into a cold 3:1 mixture of He:Ne in the
ion trap which was maintained at a nominal temperature of about 8\,K.
Typical mass spectra obtained from acetylene/\ce{CS2}-mixtures
are shown in Fig.~\ref{mass_spec_hc3sp} (top, without storing in the ion trap). Per filling cycle of the ion trap, 
some 1000 Ne$-$\hcop\ and 5000 Ne$-$\hcsp\ ions
were obtained (Fig.~\ref{mass_spec_hc3sp}, bottom). Typical trapping times were 2.5 to 3.5\,s, and irradiation (per filling cycle) was performed with $25-35$ FELIX macropulses (10\,Hz), typical pulse energies of a few up to a few tens of mJ and a Fourier-limited FWHM bandwidth of $0.5-1$~\%.
Action spectroscopy was then performed by tuning the wavenumber of FELIX light source while monitoring counts of the 
corresponding Ne-complexes of both ions (Ne$-$HC$_3$O$^+$, $m/z=73$; Ne$-$HC$_3$S$^+$, $m/z=89$). The depletion signal is power normalized prior to averaging over multiple individual scans.

To access the C–H stretching mode $\nu_1$  of \hcop\ in the 3\,$\mu$m region, 
a table-top pulsed optical parametric oscillator/amplifier (OPO/OPA, Laser Vision)
has been used  in combination with the cryogenic 22-pole trapping instrument COLTRAP~\cite{asv14} in the Cologne laboratory.
The OPO/OPA system is pumped with a pulsed 1064\,nm Nd:YAG laser operating at a repetition rate of 10\,Hz and maximum pulse energies of up to 600\,mJ. This system has been operated in an unseeded mode with a linewidth of about $0.8$\,cm$^{-1}$ (seeded operation allows a linewidth of well below $0.1$\,cm$^{-1}$) 
and pulse energies of up to 20\,mJ in the infrared range. 
The IR laser wavelength is calibrated with a wavemeter  (HighFinesse model WS-5).

\section{Quantum-chemical calculations}

Several computational investigations of 
\hcop \cite{goddard_CPL_109_170_1984,botschwina_JCP_90_4301_1989,maclagan_JCSFT_89_3325_1993,pepp_JPCA_104_5817_2000}
at various levels of theory 
%(Hartree-Fock, MP2, MP3, CEPA-1, and CCSD(T)-F12)
have been reported previously in the literature, while for
structural parameters of 
\hcsp only results at the Hartree-Fock level  \cite{maclagan_CPL_194_147_1992,liu_JPCA_101_4019_1997}
were found.
In the present study, consistently for all species under study,
quantum-chemical calculations have been performed
at the coupled-cluster singles and doubles (CCSD) level 
augmented by a perturbative treatment of triple excitations (CCSD(T))
\cite{raghavachari_chemphyslett_157_479_1989} 
together with correlation consistent (augmented) polarized valence and (augmented) polarized weighted core-valence basis sets,
i.e. 
cc-pV$X$Z, \cite{dunning_JCP_90_1007_1989} 
aug-cc-pV$X$Z,\cite{dunning_JCP_90_1007_1989, kendall_JCP_96_6796_1992,woon_JCP_98_1358_1993} 
cc-pwCV$X$Z, \cite{dunning_JCP_90_1007_1989,peterson_JCP_117_10548_2002}
and aug-cc-pwCV$X$Z \cite{dunning_JCP_90_1007_1989, kendall_JCP_96_6796_1992,woon_JCP_98_1358_1993,peterson_JCP_117_10548_2002} (with $X$=T, Q).
For basis sets denoted as cc-pV($X$+$d$)Z or aug-cc-pV($X$+$d$)Z an additional 
tight $d$ function \cite{dunning_JCP_114_9244_2001} has been added to the sulfur 
atom only, while for all other elements cc-pV($X$)Z or aug-cc-pV($X$)Z have been used, respectively. 
Equilibrium geometries have been calculated using analytic gradient techniques
\cite{watts_chemphyslett_200_1-2_1_1992}, while
harmonic frequencies have been computed using analytic second-derivative techniques
\cite{gauss_chemphyslett_276_70_1997,stanton_IntRevPhysChem_19_61_2000}.
For anharmonic computations second-order vibrational perturbation theory (VPT2) \cite{mills_alphas}   
has been employed and additional numerical differentiation of analytic second derivatives has been applied 
to obtain the third and fourth derivatives required for the application of VPT2 \cite{stanton_IntRevPhysChem_19_61_2000,stanton_JCP_108_7190_1998}.
The frozen core approximation has been indicated throughout by ``fc''.

All calculations have been carried out using 
the CFOUR program package \cite{cfour}; 
for some of the calculations the parallel version of CFOUR 
\cite{harding_JChemTheoryComput_4_64_2008} 
has been used.
% For the most part their level is identical to the geometry, so no extra single point energy computations.Those for the PES plots are specified in the following paragraph. Also, this paragraph only describes the methods and basis sets used and thats it.}

\section{Results and Discussion}

\subsection{The structures of \hcop and \hcsp and their weakly bound complexes with Ne}

\begin{figure}[H]
\centering
\includegraphics[width=14.5cm]{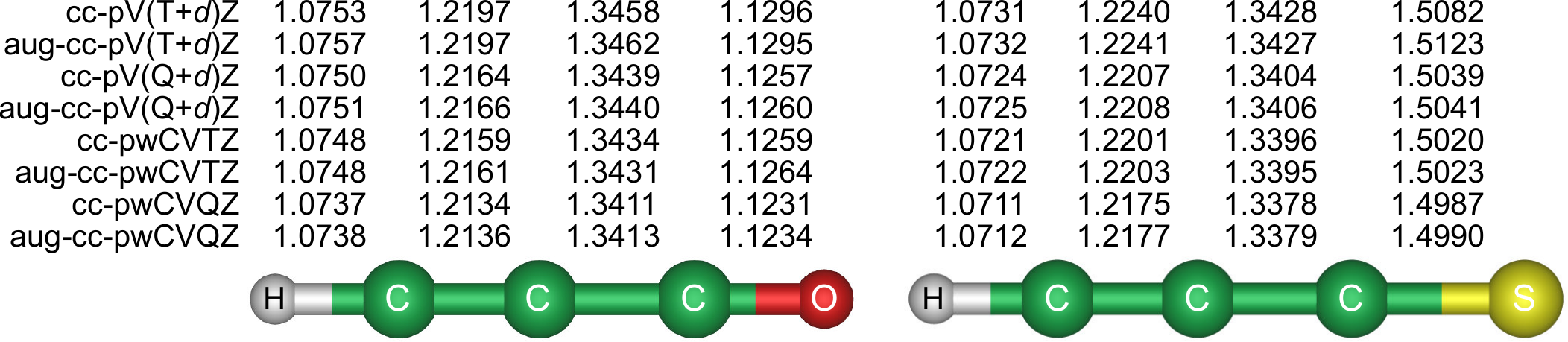}
%\subfloat[An example of an individual figure sub-caption.]{%
%\resizebox*{5cm}{!}{\includegraphics{graph1.eps}}}\hspace{5pt}
%\subfloat[A slightly shorter sub-caption.]{%
%\resizebox*{5cm}{!}{\includegraphics{graph2.eps}}}
\caption{Molecular structures of \hcop (left) and \hcsp (right) calculated at the CCSD(T) level of theory using different correlation consistent basis sets. Please note that for the cc-pV(X+$d$)Z and aug-cc-pV(X+$d$)Z basis sets the frozen-core (fc) approximation was employed and that one tight $d$-function was added only to the sulfur atom. 
Bond lengths are given in units of \AA. %\mam{I don't understand why there is results with the $d$ addition for the O-variant while the text in the caption and in the calc description says only for S. }
} \label{structures}
\end{figure}

The equilibrium structures of \hcop and \hcsp calculated at the CCSD(T) level of theory using different
correlation consistent basis sets are summarized in Fig.~\ref{structures}. The calculated bond lengths are slightly elongated for augmented basis sets relative to those calculated using the standard basis sets, while most bond lengths become shorter as the size of the basis set increases.
Over the range of basis sets employed here, the structural parameters vary by at most 6$\times$10$^{-3}$ \AA, with the exception of
the C$-$O and C$-$S distances, which vary by about 1.5$\times$10$^{-2}$ \AA.

%are displayed in Figure~\ref{structures}.\\

\begin{figure}[H]
\centering
\subfloat[Ne$-$HC$_3$O$^+$ surface.]{%
\resizebox*{14cm}{!}{\includegraphics{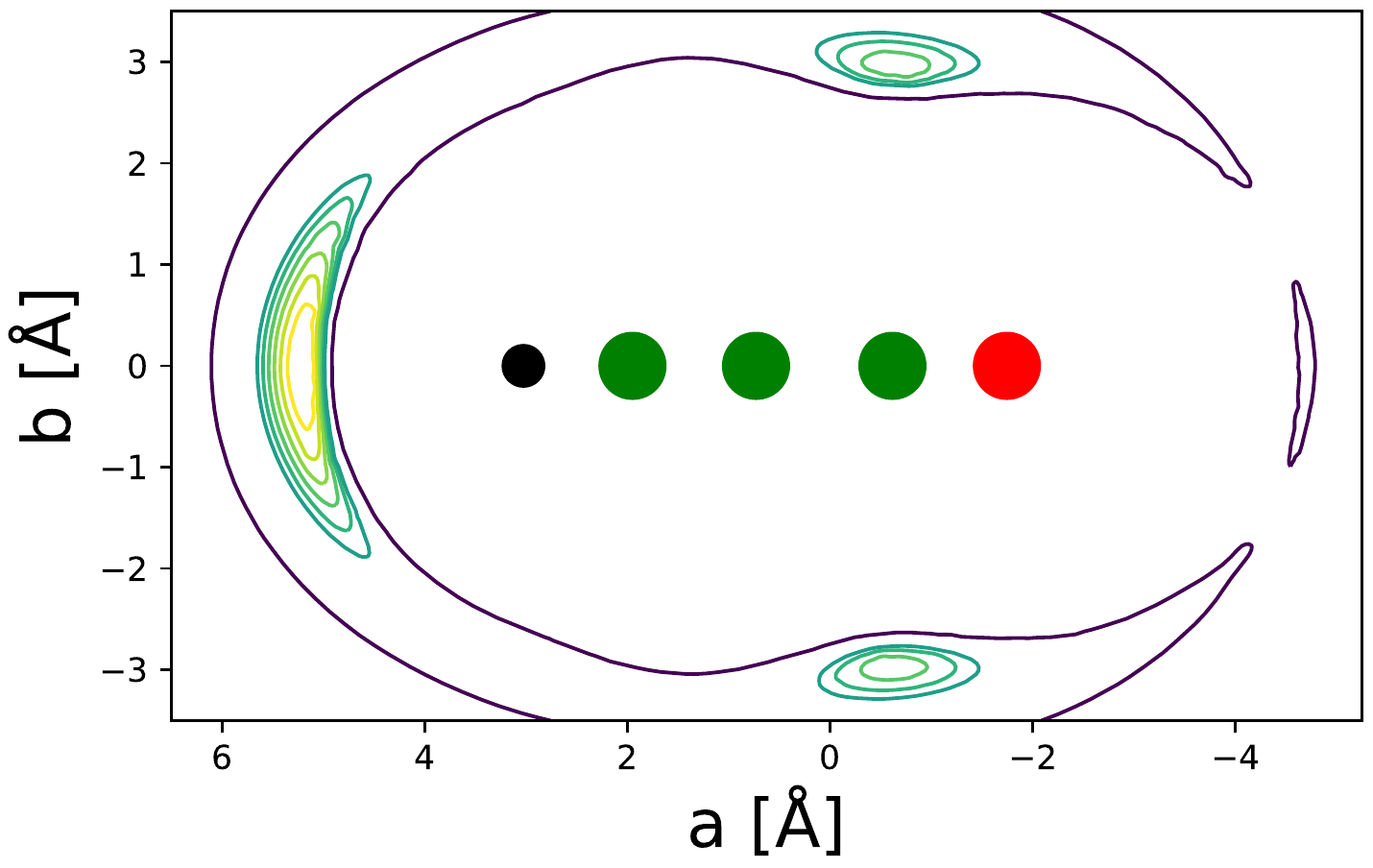}}}\hspace{5pt}
\subfloat[Ne$-$HC$_3$S$^+$ surface.]{%
\resizebox*{14cm}{!}{\includegraphics{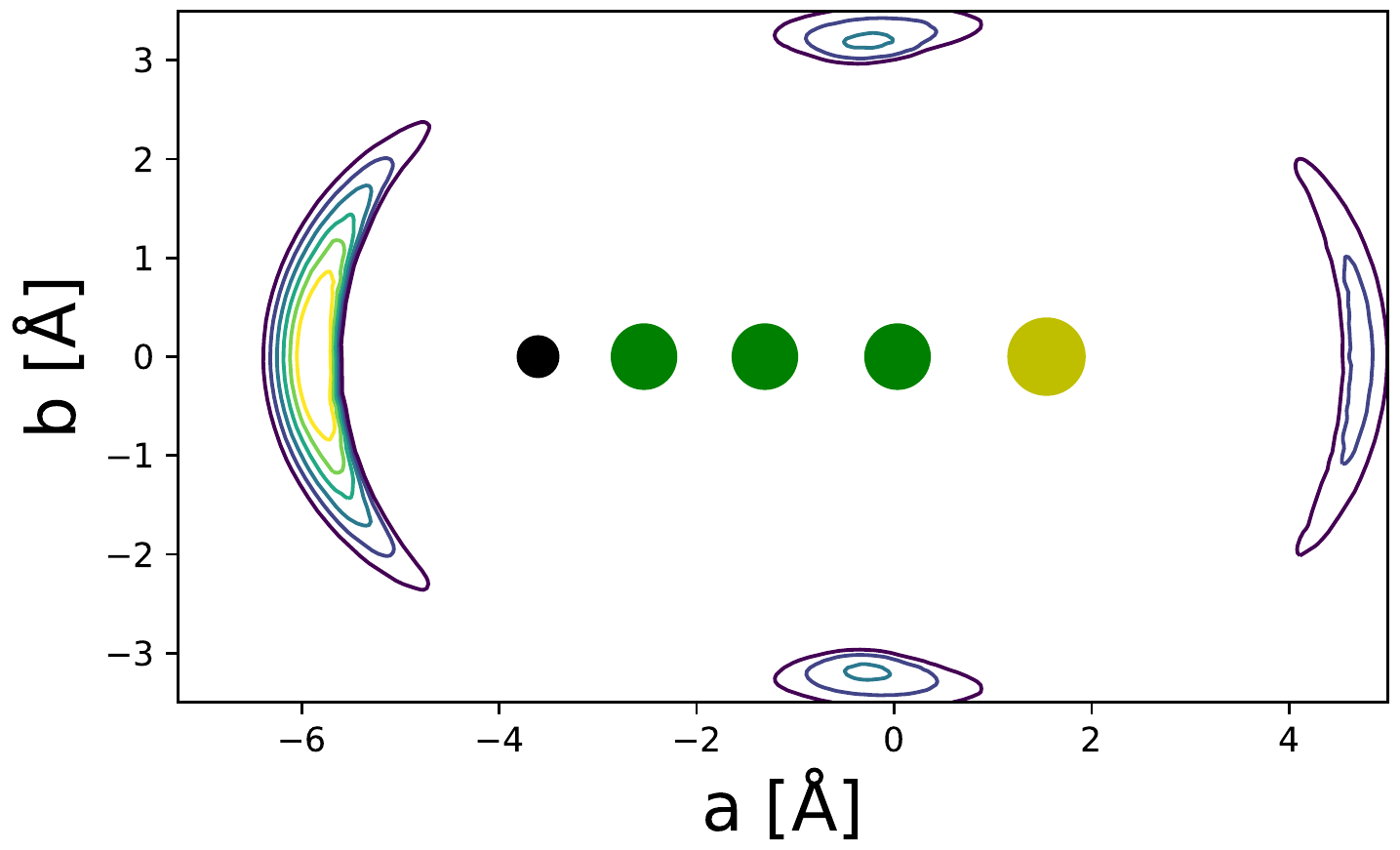}}}
\caption{Potential energy landscape with respect to the position of the Ne atom in the two  Ne-HC$_3$X$^+$ complexes calculated at the fc-CCSD(T)/aug-cc-pV(T+$d$)Z level of theory.
Atom color code: oxygen (red), sulfur (yellow), carbon (green), hydrogen (black). Contours cover the interval [0.05,0.30] kcal/mol in steps of 0.05\,kcal/mol above the global minimum. In case of \hcop\ one additional contour is located at 0.615\,kcal/mol to visualize the shallow minimum of the \ce{HC3O+}$-$Ne arrangement.} \label{peshcop}
\end{figure}

As a compromise between accuracy and computational demand, the Ne-complexes of \hcop and \hcsp were calculated at the 
fc-CCSD(T)/aug-cc-pV(T+$d$)Z level of theory. 
To determine  the most energetically favorable locations of the Ne atom with respect to the \hcop and \hcsp chains, i.e., to identify local minima on the Ne$-$\hcop and Ne$-$\hcsp potential energy surfaces,
the fc-CCSD(T)/aug-cc-pV(T+$d$)Z equilibrium structures in Fig.~\ref{structures} were kept fixed 
and the position of the Ne atom was varied on a 11$\times$7\,\AA$^2$\ grid using a spacing of 0.25~\AA\ 
and distances ranging from some 1.5 to 3.5\,\AA\ about the \hcop and \hcsp\ chains.
For each of the some 500 grid points for both surfaces single-point energy calculations were performed,
the results of which are graphically represented in Fig.~\ref{peshcop}.  
This approach led to the identification of three minima for each complex:
a H-bound linear structure, which is the global minimum for both complexes (e.g., similar to that
of the Ne$-$\ce{HCO+} cluster  \cite{nizkorodov_JCP_105_1770_1996}),
followed by non-linear (T-shaped) variants. A third, weakly-bound minimum exists in which the Ne atom is located adjacent to the O and S atoms in a linear arrangement.

In a second approach, the structural optimizations of all 
minimum energy configurations were fully relaxed.  
As indicated in the structural 
parameters collected in Table~\ref{calcdisthc3o+}, the perturbation of Ne on both \hcop and \hcsp 
is small, with variations on the order of $10^{-3}$\,\AA .
%of bond distances in the Ne-HC$_3$O$^+$ and Ne-HC$_3$S$^+$ complexes as seen in Table~\ref{calcdisthc3o+}
%is quite small. 
The Ne bond dissociation energies $D_e$  differ by at most 0.6\,kcal mol$^{-1}$ in the minimum configurations %effects 
identified,  amounting to 
1.0\,kcal mol$^{-1}$, 0.9\,kcal mol$^{-1}$, and 0.4\,kcal mol$^{-1}$ for Ne$-$\hcop and
0.9\,kcal mol$^{-1}$, 0.7\,kcal mol$^{-1}$, and 0.6\,kcal mol$^{-1}$ for Ne$-$\hcsp .
When rough estimates of the zero-point vibrational effects on the Ne bond dissociation energies ($D_0$) are included, the H-bound and T-shaped forms are essentially isoenergetic for \hcop, while the O-bound form is slightly less stable by about 0.5\,kcal mol$^{-1}$.
For \hcsp, the energy difference between the three structures is more or less negligble.

%The 5.6\,kJ mol$^{-1}$ (=468\,\wn ) bond dissociation energies calculated here corresponds very well
%with the value derived for Ne$-$\ce{HCO+} 

%that also summarize the bond dissociation energies of neon.
%at the fc-CCSD(T)/aug-cc-pV(T+d)Z level of theory.

%As expected the qualitative structures are 
%very similar and after full relaxation at the fc-CCSD(T)/aug-cc-pV(T+d)Z level of theory the 
%corresponding local 
%minima are obtained see Table~\ref{calcdisthc3o+}.

\begin{table}[H]
\tbl{Variation of bond distances of \hcop and \hcsp with respect to Ne tagging
as well as Ne bond dissociation energies $D$. $D_e$ denotes the electronic contribution at the equilibrium geometry, 
while $D_0$ denotes $D_e$ augmented by a harmonic estimate of the zero-point vibrational contribution.
All results have been obtained at the fc-CCSD(T)/aug-cc-pV(T+$d$)Z level of theory.
Bond lengths given in units of \AA, Ne bond dissociation energies in kcal mol$^{-1}$.}
{\begin{tabular}{lrrrr|rr} 
\toprule
Species                                     & R(CH)  &  R(CC) & R(CC)  & R(CO/S) & $D_e$ & $D_0$\\
\hline 
HC$_3$O$^+$                                 & 1.0757 & 1.2197 & 1.3462 & 1.1295 & -     & -\\
Ne-HC$_3$O$^+$ (lin.)                       & 1.0763 & 1.2199 & 1.3459 & 1.1297 &  1.0  &  0.7\\
Ne-HC$_3$O$^+$ (nonlin.)\textsuperscript{a} & 1.0756 & 1.2195 & 1.3463 & 1.1294 &  0.9  &  0.7\\
HC$_3$O$^+$-Ne (lin.)                       & 1.0757 & 1.2197 & 1.3463 & 1.1295 &  0.4  &  0.3\\
\hline 
HC$_3$S$^+$                                 & 1.0732 & 1.2241 & 1.3427 & 1.5123 & -     & -\\
Ne-HC$_3$S$^+$ (lin.)                       & 1.0735 & 1.2243 & 1.3423 & 1.5125 &  0.9  & 0.6\\
Ne-HC$_3$S$^+$ (nonlin.)\textsuperscript{a} & 1.0731 & 1.2240 & 1.3427 & 1.5122 &  0.7  & 0.6\\
HC$_3$S$^+$-Ne (lin.)                       & 1.0732 & 1.2241 & 1.3429 & 1.5124 &  0.6  & 0.5\\
 \bottomrule
\end{tabular}}
\tabnote{\textsuperscript{a} Deviations from linearity in the bond angles of the non-linear (T-shaped) forms are found within a maximum of 1 degree, see Appendix \ref{appendix_a}.}
\label{calcdisthc3o+}
\end{table}

\subsection{The influence of Ne-tagging on the vibrational spectra of \hcop and \hcsp}

As linear pentaatomic species, \hcop\ and \hcsp\ both possess seven fundamental vibrational
modes, four stretching ($\sigma$) and three doubly degenerate ($\pi$) bending modes.
%While neon is bound very weakly to HC$_3$X$^+$ in all three complexed identified, 
%the variation in the structural parameters is as expected as well very small.
To further investigate the perturbation caused by a single Ne atom on their vibrational spectra,% of HC$_3X^+$,
harmonic vibrational wavenumbers of \hcop, \hcsp,  and their corresponding complexes with Ne were calculated
at the fc-CCSD(T)/aug-cc-pV(T+$d$)Z level of theory. %have been evaluated 
As shown in Table~\ref{calcvibhc3o+},
%The harmonic frequencies shown in Table~\ref{calcvibhc3o+} for \hcop, \hcsp  and the corresponding complexes with Ne 
these values 
differ by a only few \wn. One notable exception are the C$-$C$-$H-bending modes $\omega_5$ of the 
H-bound  Ne$-$\hcop and Ne$-$\hcsp linear isomers. Here Ne-tagging results in a 25 to 30\,\wn blueshift of the modes.  
Nevertheless,  no large shifts are expected for any of the vibrational modes in \hcop and \hcsp upon Ne-tagging.
%with the exception $\nu_5$($\pi$) of  \hcop and \hcsp, respectively,
%which is the ben

\begin{table}[H]
\tbl{Harmonic frequencies (in \wn) of \hcop, \hcsp,  and the corresponding complexes with Ne 
evaluated at the fc-CCSD(T)/aug-cc-pV(T+$d$)Z level of theory.}
{\begin{tabular}{lrrrr|rrrr} 
\toprule
Mode\textsuperscript{a} & HC$_3$O$^+$   &  Ne--HC$_3$O$^+$ & Ne--HC$_3$O$^+$ & HC$_3$O$^+-$Ne & HC$_3$S$^+$   &  Ne--HC$_3$S$^+$ & Ne--HC$_3$S$^+$ & HC$_3$S$^+-$Ne\\
     &          &  \multicolumn{1}{c}{linear\textsuperscript{a}}          & \multicolumn{1}{c}{nonlinear\textsuperscript{a}}      & \multicolumn{1}{c|}{linear\textsuperscript{a}}       &               & \multicolumn{1}{c}{linear\textsuperscript{a}}          & \multicolumn{1}{c}{nonlinear\textsuperscript{a}}      & \multicolumn{1}{c}{linear\textsuperscript{a}}        \\      
\hline 
$\omega_1$                      & 3351 &  3347  & 3352       & 3351 & 3370 & 3370   & 3371     & 3370  \\
$\omega_2$                      & 2346 &  2345  & 2347       & 2346 & 2133 & 2132   & 2134     & 2133  \\
$\omega_3$                      & 2094 &  2093  & 2095       & 2094 & 1634 & 1634   & 1634     & 1633  \\
$\omega_4$                      & 916  &  918   & 916        &  916 &  730 &  731    &  730     &  730  \\
$\omega_5$\textsuperscript{c,d}   & 775  &  806   & 774/774  &  774 &  733 &  758    &  732/732 &  732  \\
%$\nu_5$($\pi$)                 & 775  &  806   & 774      &  774   &  733 &  758    &  732     &  732  \\
$\omega_6$\textsuperscript{c,d}   & 546  &  547   & 546/544  &  547 &  472 &  473    &  471/471 &  473  \\
%$\omega_6$                     & 546  &  547   & 544      &  547   &  472 &  473    &  471     &  473  \\
$\omega_7$\textsuperscript{c,d}   & 162  &  168   & 163/161  &  162 &  176 &  183    &  177/176 &  177  \\
%$\nu_7$($\pi$)                 & 162  &  168   & 161       &  162  &  176 &  183    &  176     &  177  \\
$\omega_8$\textsuperscript{b}   & -    &  75    & 61         &  48  &    - &   67    &   48     &   48  \\
$\omega_9$\textsuperscript{b,c} & -    &  33    & 31         &  10  &    - &   29    &   22     &    9  \\
%$\nu_9$($\sigma$)              & -    &  33    &-          &  10  &     - &   29    &   -      &    9  \\
 \bottomrule
\end{tabular}}
\tabnote{\textsuperscript{a} Mode index borrowed from untagged \hcop and \hcsp for the sake of comparability.}
\tabnote{\textsuperscript{b} Extra low-frequency vibrational modes introduced in the Ne--\hcop and Ne--\hcsp complexes, arbitrary mode index.}
\tabnote{\textsuperscript{c} Doubly degenerate bending mode in linear species.}
\tabnote{\textsuperscript{d} Degeneracy is lifted in the non-linear forms.}
%\tabnote{\textsuperscript{d} Mode designation between $\omega_4$ and $\omega_5$ modes flipped to conform to designation used for \hcsp in section\ref{section-hc3sp}.}
\label{calcvibhc3o+}
\end{table}

\subsection{Experimental spectra of Ne$-$\hcop and Ne$-$\hcsp}

The IR results for \hcop and \hcsp are summarized here. It should again be emphasized  
that the present IRPD scheme traces the weakly bound complexes
with Ne rather than the bare ions themselves, 
but, as demonstrated in the previous section, the perturbation of the Ne atom on the vibrational fundamentals of the ions is very small. i.e.~a few \wn at most. 
The anharmonic vibrational 
force fields guiding spectroscopic assignment and analysis were calculated
at the fc-CCSD(T)/cc-pVTZ (\hcop ) and fc-CCSD(T)/cc-pV(T+$d$)Z 
(\hcsp) levels. In an attempt to further improve the vibrational wavenumbers
calculated for \hcop , scaling factors 
derived from a comparison of the corresponding experimental and calculated values of
isoelectronic \ce{HC3N} were employed (Table \ref{vibfundhcop}).
These empirical corrections, however, were found to be very small ($\le 9$\,\wn ) 
and do not affect any spectroscopic assignment. They are given here nonetheless for the sake
of completeness.

\begin{table}[H]
\tbl{Fundamental vibrational wavenumbers of HC$_3$N and \hcop (in \wn ) and IR band intensities of \hcop (km/mol).}
{\begin{tabular}{lrrr|rrrrrr} \toprule
 & \multicolumn{3}{c}{HC$_3$N} & \multicolumn{5}{c}{\hcop} \\ \cmidrule{2-4} \cmidrule{5-9}
%  & \multicolumn{2}{c}{Calculation} &  \\ \cmidrule{2-3} \cmidrule{4-6}
 Mode    & Harm\textsuperscript{a} & Anharm\textsuperscript{a} &  Exp\textsuperscript{b}   & Harm\textsuperscript{a} & Anharm\textsuperscript{a} & BE\textsuperscript{c} & Exp\textsuperscript{d} & Int\textsuperscript{a,e} \\ \midrule
 $\nu_1$($\sigma$) C$-$H stretch        & 3459 & 3325 & 3327   & 3362 & 3229 & 3231 &  3232      & 96 \\
 $\nu_2$($\sigma$) C$\equiv$N/O stretch & 2312 & 2271 & 2274   & 2353 & 2313 & 2316 &  2313      & 782 \\
 $\nu_3$($\sigma$) C$\equiv$C stretch   & 2104 & 2069 & 2079   & 2097 & 2064 & 2074 &  $\cdots$  & 104 \\
 $\nu_4$($\sigma$) C$-$C stretch        &  875 &  853 &  862   &  918 &  901 &  911 &   906      &   7 \\
 $\nu_5$($\pi$) C$-$C$-$H bending       &  663 &  650 &  663.2 &  773 &  758 &  773 &   764      &  54 \\
 $\nu_6$($\pi$) C$-$C$-$N/O bending     &  498 &  486 &  498.8 &  552 &  547 &  558 &   555      &  54 \\
 $\nu_7$($\pi$) C$-$C$-$C bending       &  227 &  217 &  222.4 &  181 &  166 &  169 &  $\cdots$  &   6 \\
% Alpha\textsuperscript{a} & A1 & A2 & A3 & A4 & A5 & A6 \\
% Beta & B2 & B2 & B3 & B4 & B5 & B6 \\
% Gamma & C2 & C2 & C3 & C4 & C5 & C6 \\ 
\bottomrule
\end{tabular}}
\tabnote{\textsuperscript{a} fc-CCSD(T)/cc-pVTZ.}
\tabnote{\textsuperscript{b} Experimental vibrational wavenumbers from Ref. \cite{Jolly_JMS_242_46_2007}.}
\tabnote{\textsuperscript{c} Best estimate (BE) value: Anharmonic vibrational wavenumber of \hcop scaled
by the ratio of Exp/Anharmonic wavenumbers of \ce{HC3N}.}
\tabnote{\textsuperscript{d} See Figure \ref{hc3op_FELIX_spec}.}
\tabnote{\textsuperscript{e} IR intensities obtainted via VPT2.}
\label{vibfundhcop}
\end{table}

\subsubsection{FELIX and OPO spectra of Ne$-$\ce{HC3O+}}

\begin{figure}[H]
\centering
\includegraphics[width=14.5cm]{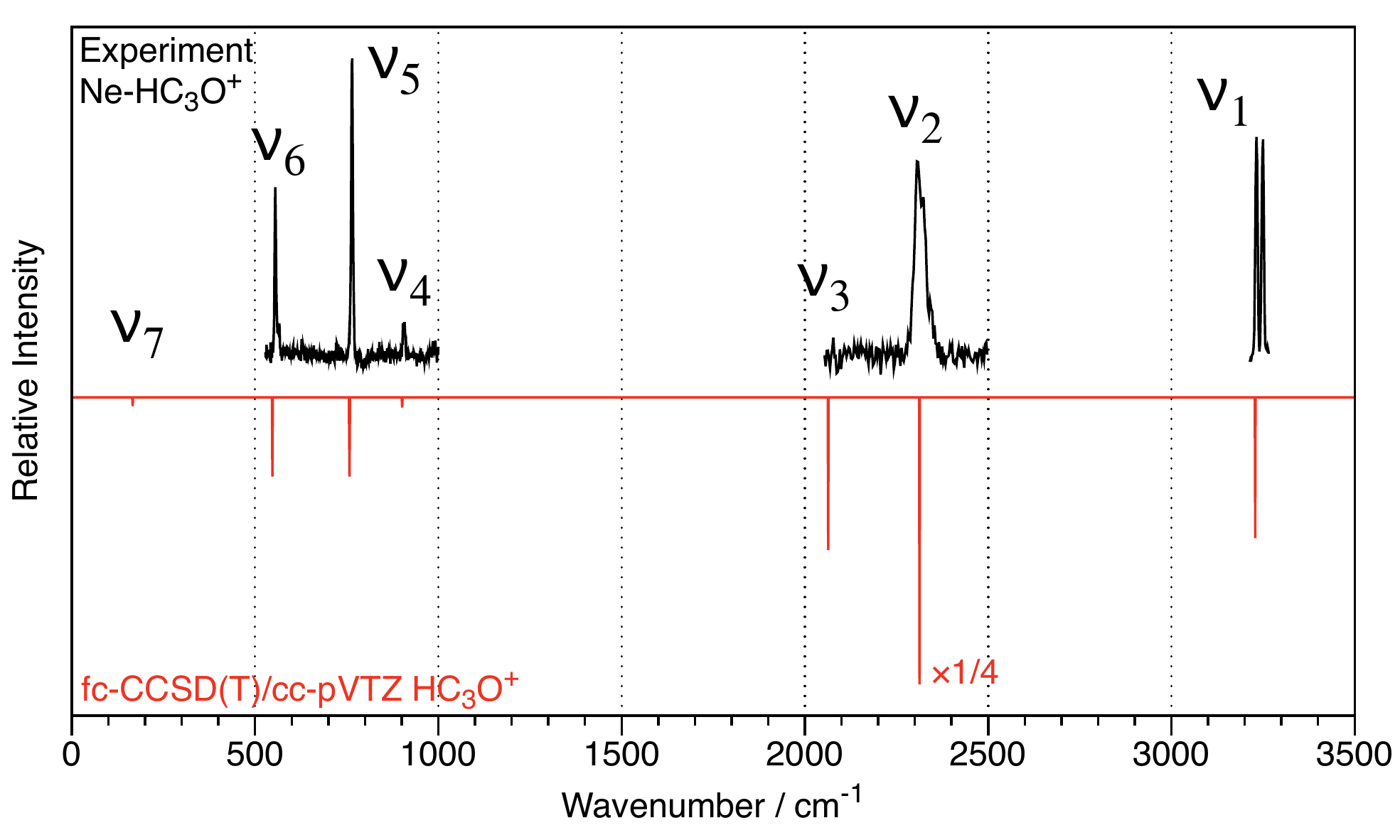}
%\subfloat[An example of an individual figure sub-caption.]{%
%\resizebox*{5cm}{!}{\includegraphics{graph1.eps}}}\hspace{5pt}
%\subfloat[A slightly shorter sub-caption.]{%
%\resizebox*{5cm}{!}{\includegraphics{graph2.eps}}}
\caption{The vibrational spectrum of Ne$-$\ce{HC3O+} as observed employing IRPD with 
FELIX between 500 and 2500\,\wn and an optical parametric oscillator around 3200\,\wn 
in the Cologne 
laboratory (top, black). Each peak in the experimental spectrum 
corresponds to a power-normalized relative depletion of the Ne$-$\hcop counts upon excitation of a vibrational transition.
For comparison, the results from an 
VPT2 calculation of the bare ion performed at the 
fc-CCSD(T)/cc-pVTZ level of theory are shown as red sticks. For better visibility of all calculated fundamentals in the stick simulation, the intensity of the $\nu_2$ mode has been multiplied by 1/4. 
At the predicted location of the $\nu_1$ mode, the experimental spectrum shows a doublet separated by about 17\,\wn , see text. 
} \label{hc3op_FELIX_spec}
\end{figure}

The FELIX IRPD spectra of Ne$-$\hcop are shown in Figure \ref{hc3op_FELIX_spec}.
For comparison, a simulation based on results of an anharmonic VPT2 calculation at the fc-CCSD(T)/cc-pVTZ level of the bare \hcop ion is also presented. Wavenumbers of the experimental vibrational bands and those predicted theoretically are compared in Table \ref{vibfundhcop}.
Experimentally, two regions were covered with FELIX: 530 to 1000\,\wn and 
2060 to 2500\,\wn . Four prominent features, all of which coincide
virtually quantitatively with the calculated vibrational fundamentals of \hcop, were observed in these two regions. The features are identified as the $\nu_2$ fundamental (C$-$O stretch) at 2313~\wn (best estimate value at
2316\,\wn ), $\nu_4$ (C$-$C stretch) at 906\,\wn (best estimate value 911\,\wn ),
$\nu_5$  (C$-$C$-$H bending) at 764~\wn (best estimate 773\,\wn ), and $\nu_6$ (C$-$C$-$O bending) at 555\,\wn (best estimate at 558\,\wn ). 
Unfortunately, the $\nu_3$ mode (C$\equiv$C stretch) predicted at 2074\,\wn, was not detected. Presumably this band lies just outside the region accessible with FELIX during our observing run. As shown in section \ref{section-hc3sp}, the corresponding mode  of Ne$-$\hcsp is  rather prominent.

%No clear feature was detected for the $\nu_3$ mode (C$\equiv$C stretch) predicted at 2074~\wn .
%Presumably, this band is located just outside the region that was accessible with FELIX during the observing run.
%\begin{table}[H]
%\tbl{Vibrational fundamentals of HC$_3$N and \hcop .}
%{\begin{tabular}{lrrrrrc} \toprule
% & \multicolumn{2}{c}{HC$_3$N} & \multicolumn{3}{c}{\hcop} \\ \cmidrule{2-3} \cmidrule{4-6}
% Mode    & Calc &  Exp\textsuperscript{a}  & Calc & scaled & Exp  \\ \midrule
% $\nu_1$($\sigma$), C$-$H stretching       & 3325 & 3327   & 3229 & 3231 &  3232/3249 \\
% $\nu_2$($\sigma$), C$\equiv$O stretching  & 2271 & 2274   & 2313 & 2316 &  2313      \\
% $\nu_3$($\sigma$), C$\equiv$C stretching  & 2069 & 2079   & 2064 & 2074 & (2071)     \\
% $\nu_4$($\sigma$), C$-$C stretching       &  853 &  862   &  901 &  911 &   906      \\
% $\nu_5$($\pi$), CCH-bending               &  650 &  663.2 &  758 &  773 &   764      \\
% $\nu_6$($\pi$), CCO-bending               &  486 &  498.8 &  547 &  558 &   555      \\
% $\nu_7$($\pi$), CCC-bending               &  218 &  222.4 &  166 &  169 &  $\cdots$          \\
%% Alpha\textsuperscript{a} & A1 & A2 & A3 & A4 & A5 & A6 \\
%% Beta & B2 & B2 & B3 & B4 & B5 & B6 \\
%% Gamma & C2 & C2 & C3 & C4 & C5 & C6 \\ 
%\bottomrule
%\end{tabular}}
%\tabnote{\textsuperscript{a} Ref. \cite{Jolly_JMS_242_46_2007}}
%\label{vibfundhcop}
%\end{table}
%\subsection{The $\nu_1$ mode of Ne$-$\ce{HC3O+}}
The $\nu_1$ mode around 3200\,\wn was detected in the Cologne laboratory using a commercial
pulsed OPO system. Based on the empirically-scaled predictions from Table \ref{vibfundhcop}, our best estimate was  3231\,\wn for this mode,
 and indeed it was found very close by, at 3232\,\wn . % (Figure \ref{hc3op_OPO_spec}). 
Scanning further up in energy revealed a second band of similar intensity at 3249\,\wn. The origin of this second band 
is not entirely clear, since only one band is expected for the bare ion. 
Also, a combination of the $\nu_2$ and $\nu_4$ modes seems unlikely, which would be expected a few tens of \,\wn\ below $\nu_1$.
As the calculated Ne-induced shifts of this band for any of the three different complexes (see Table \ref{calcvibhc3o+})
are much smaller than the splitting
observed, the possibility of a second structural isomer seems unlikely but cannot be ruled out.
However, combination modes
involving the Ne atom might need to be considered further. 
Such vibrational ``tag-satellites" have been observed previously~\cite{oku85,kohguchi_JCP_148_144303_2018}, 
albeit at much less intensity.
Assuming that the quantum-chemical calculations are correct, and the complex being probed is the linear H-bound form (global minimum), the
Ne$-$H$-$C bending mode would be the energetically lowest vibrational fundamental in the Ne$-$\hcop complex.
While anharmonic force field calculations of weakly-bound molecules employing VPT2 are intrinsically difficult,
trial calculations performed here at the fc-CCSD(T)/aug-cc-pVTZ level %(Table \ref{calcvibhc3o+anh}) 
predict this mode has a vibrational frequency of roughly 10\,\wn. Hence, it is possible that the second mode is in fact a combination mode of C$-$H stretching and Ne$-$H$-$C bending. 

Additional measurements at high spectral resolution may help to resolve this issue and others in the IR spectrum of Ne$-$\ce{HC3O+}.  For example, an analysis of the underlying rotational structure for the $\nu_1$ mode  would likely clarify the carrier
of these spectroscopic features, and in doing so provide a stringent test of the present spectroscopic assignment.
Rotational resolution of the rather broad $\nu_2$ mode observed with FELIX would also be key to understanding 
possible causes for its width, such as lifetime broadening or the more speculative explanation that it arises from tag-satellites from the Ne tag.
%As the action spectroscopy performed here 

%As can be seen from Figure \ref{hc3op_FELIX_spec}, the $\nu_2$ mode
%is found comparable broad. The 

%\begin{figure}[H]
%\centering
%\includegraphics[width=10.5cm]{hccco+Ne-opo_pavol_032020_paper.pdf}
%\caption{The $\nu_1$ vibrational mode region of Ne$-$\ce{HC3O+} cluster observed with the Cologne OPO. 
%} \label{hc3op_OPO_spec}
%\end{figure}

\subsubsection{Ne$-$\ce{HC3S+} observed with FELIX}
\label{section-hc3sp}

\begin{figure}[H]
\centering
\includegraphics[width=14.5cm]{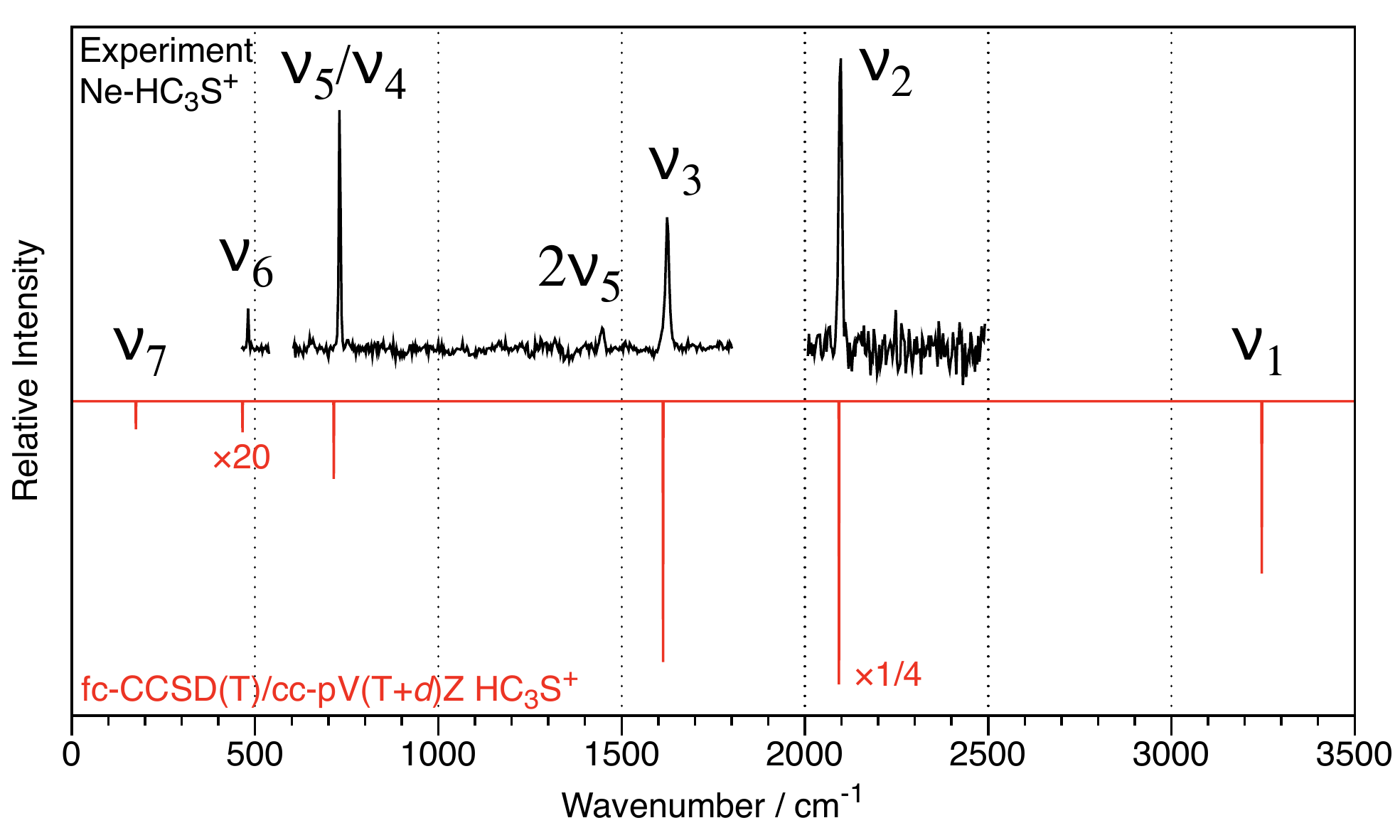}
%\subfloat[An example of an individual figure sub-caption.]{%
%\resizebox*{5cm}{!}{\includegraphics{graph1.eps}}}\hspace{5pt}
%\subfloat[A slightly shorter sub-caption.]{%
%\resizebox*{5cm}{!}{\includegraphics{graph2.eps}}}
\caption{The vibrational spectrum of Ne$-$\ce{HC3S+} as observed with 
FELIX between 480 and 2500\,\wn (top, black). Each peak in the experimental spectrum 
corresponds to a power-normalized relative depletion of the Ne$-$\hcsp counts upon excitation of a vibrational transition. For comparison, the results from an 
VPT2 calculation of the bare ion performed at the 
fc-CCSD(T)/cc-pV(T+$d$)Zd level of theory are shown as red sticks. For better visibility of all calculated fundamentals in the simulation, the intensity of the $\nu_2$ band has been multiplied by 1/4. The calculated intensity of the $\nu_6$ mode was multiplied by a factor of 20 to be visible in the stick spectrum. The weakest feature in the experimental spectrum at 1446~\wn is not from a vibrational fundamental but most likely from an overtone, the 2$\nu_5$ mode, see text.
} \label{hc3sp_spec}
\end{figure}

The FELIX spectra of Ne$-$\hcsp are shown in Fig.~\ref{hc3sp_spec} along with a simulation based on results from an anharmonic VPT2 calculation at the fc-CCSD(T)/cc-pV(T+$d$)Z level of the bare ion. As the vibrational spectrum
of isoelectronic \ce{HC3P} has not yet been studied experimentally, no empirical scaling was possible.
Experimentally, three regions were covered with FELIX: 
460 to 540\,\wn , 600 to 1800\,\wn ,
as well as 2010 to 2490\,\wn .
In total, four prominent features are observed; the frequencies of each agree
very well with vibrational fundamentals calculated for \hcsp : $\nu_2$ observed at
2097\,\wn  (calculated at 2093\,\wn; see Table~\ref{vibfundhcsp}),
$\nu_3$ at
1624\,\wn (calculated at 1613\,\wn ), %1624\,\wn (calculated at 1613\,\wn ),
$\nu_4$ and $\nu_5$ most likely overlapped to one feature observed at 731\,\wn (calculated at 719 and 715\,\wn , respectively), and finally the $\nu_6$ mode at 481\,\wn 
(calculated at 466\,\wn ). The C$-$H stretching ($\nu_1$) and C$-$C$-$C ($\nu_7$) bending modes calculated
at 3247\,\wn and 175\,\wn lie outside our our measurement range, and were not observed experimentally here. The weakest feature
in the Ne$-$\hcsp spectrum is observed at 1446\,\wn . This feature is not from a vibrational
fundamental, however, it is located close to two times the wavenumber
of the $\nu_4$/$\nu_5$ fundamental feature observed at 731\,\wn and hence very likely corresponds to an overtone/combination mode. While the intensities observed in action spectra are not strictly 
related to spectroscopic band intensities, they are often found to be in qualitative agreement.
From this viewpoint, it can be assumed that both the 731\,\wn and 1446\,\wn features
are dominated by the contribution from the $\nu_5$ mode whose IR intensity in the bare ion
is calculated to be 500 times higher than that of $\nu_4$ (Table \ref{vibfundhcsp}).

%the theoretical prediction very well. The C$\equiv$C stretch $\nu_3$ observed at 2091\,\wn,  

%The experimental spectra show four prominent spectroscopic features, all of which coincide
%virtually quantitatively with calculated vibrational fundamentals of \hcop .
%These are the
%modes $\nu_2$ (C$-$O stretch) located at 2313\wn , $\nu_4$ (C$-$C stretch) at 906\,\wn ,
%$\nu_5$ (CCH bending) at 765\wn , and $\nu_6$ (CCO bending) at 555\,\wn . No clear feature was detected for the $\nu_3$ (C$\equiv$C stretch) predicted at 2074\wn ,
%Presumably, this band is located just outside the region that was accessible with FELIX
%during the \ce{HC3O+} observing campaign.

\begin{table}[H]
\tbl{Fundamental vibrational wavenumbers of \hcsp (in \wn ) and IR band intensities (km/mol).}
{\begin{tabular}{lrrrr} \toprule
 & \multicolumn{4}{c}{\hcsp} \\ \cmidrule{2-5} 
 Mode                         & Harm\textsuperscript{a} &  Anharm\textsuperscript{a} &  Exp\textsuperscript{b}      & Int\textsuperscript{a,e}  \\ \midrule
 $\nu_1$($\sigma$) C$-$H stretch      & 3381 &  3247 & $\cdots$  &  111  \\
 $\nu_2$($\sigma$) C$\equiv$C stretch & 2137 &  2093 & 2097      &  730   \\
 $\nu_3$($\sigma$) C$-$S stretch      & 1636 &  1613 & 1624      &  168   \\
 %$2\nu_5$($\sigma$)C$-$C stretch      &      &  1423 & 1446      &        \\
 $\nu_4$($\sigma$) C$-$C stretch      &  729 &   719 &  731      &   0.1  \\
 $\nu_5$($\pi$)    C$-$C$-$H bending  &  738 &   715 &  731      &   50   \\
 $\nu_6$($\pi$)    C$-$C$-$S bending  &  485 &   466 &  481      &    1   \\
 $\nu_7$($\pi$)    C$-$C$-$C bending  &  185 &   175 &  $\cdots$ &   18   \\
% Alpha\textsuperscript{a} & A1 & A2 & A3 & A4 & A5 & A6 \\
% Beta & B2 & B2 & B3 & B4 & B5 & B6 \\
% Gamma & C2 & C2 & C3 & C4 & C5 & C6 \\ 
\bottomrule
\end{tabular}}
\tabnote{\textsuperscript{a} fc-CCSD(T)/cc-pVTZ.}
%\tabnote{\textsuperscript{b} Experimental vibrational wavenumbers from Ref. \cite{Jolly_JMS_242_46_2007}.}
%\tabnote{\textsuperscript{c} Anharmonic vibrational wavenumber of \hcop scaled
%by the ratio of Exp/Anharmonic wavenumbers of \ce{HC3N}.}
\tabnote{\textsuperscript{b} See Figure \ref{hc3sp_spec}.}
\tabnote{\textsuperscript{e} IR intensities obtainted via VPT2.}
%\tabnote{\textsuperscript{a} Ref. }
\label{vibfundhcsp}
\end{table}

\section{Conclusions and Prospects}

The present study marks the first spectroscopic detection of two fundamental molecular ions
of potential astronomical relevance, \hcop\ and \hcsp . The detections were made possible by using very sensitive IR action spectroscopic methods in combination with ion traps, an approach that can likely be adapted to study similar and even (much) more complex species at high spectral resolution. Owing to mass selection
prior to spectroscopic interrogation in the ion trap, the resulting spectra are free from contamination and can be readily assigned using high-level quantum-chemical calculations of the bare ion since coupling of the Ne tag to the ion is very weak. 
%The calculations also reveal that the method of Ne-tagging employed here has a very small effect on the location of the vibrational bands compared to the bare ion. 

While the molecular carriers of the vibrational features studied here, \hcop and
\hcsp , have been established
beyond any reasonable doubt, the present measurements lack the rotational information that is required for more detailed spectroscopic analysis or to undertake a 
radio astronomical search. Nevertheless, the present work provides a firm foundation for subsequent analysis in this and other wavelength regions.  %Future efforts should be undertaken towards obtaining this information. 
For example, now that the positions of several vibrational bands of \ce{HC3O+} and
\ce{HC3S+} are known, the use of narrow-linewidth continuous-wave IR sources such as quantum cascade lasers or optical parametric oscillators may permit the observation of selected bands
at very high spectral resolution. Recent examples of such studies in the Cologne laboratory include the observation of CH$^+$ and \ce{CD2H+} using laser induced reaction (LIR)
or laser induced inhibition of complex growth (LIICG) schemes \cite{dom18,jusko_JMS_332_59_2017}, which provide spectroscopic information on the bare ions.
Analogous higher-resolution studies of the Ne-ion complexes described here would allow to determine their precise geometrical structure and resolve the issue surrounding the assignment of the peculiar second band observed in the vicinity of the $\nu_1$ band of \hcop. 

In addition to any studies in the infrared, rotational spectra may be observed directly using the method of rotational state-specific attachment of rare gas atoms (ROSAA) to cations  
at very low temperatures \cite{brunken_JMS_332_67_2017}, a method that has been used recently with excellent success to obtain the pure rotational spectra of fundamental ions \cite[see, e.g.,][]{dom17,thorwirth_ApJ_2019,markus_2019}.
The rotational spectra of two species isoelectronic with \hcop and
\hcsp -- \ce{HC3N} and \ce{HC3P} -- have been studied very accurately \cite[see Refs.][and references therein]{thorwirth_JMolSpectrosc_204_133_2000,bizzocchi_JMS_110_205_2001}, 
and may be used to calibrate the computational results of the two ions obtained here.
For \ce{HC3N}, the ground state rotational constant $B_0$ from a CCSD(T)/cc-pwCVQZ calculation
and CCSD(T)/cc-pVTZ zero point vibrational corrections is 4542.7\,MHz and hence deviates
by only 0.14\% from the experimental value of 4549.059\,MHz. Using the ratio $B_\mathrm{exp}/B_\mathrm{calc}$ of \ce{HC3N} for scaling $B_0=4454.2$\,MHz of \hcop calculated at the same levels of theory
yields a best estimate value of 4460.4\,MHz (for $B_0$ of \hcop ). For \ce{HC3P}, 
this treatments yields $B_{0,\,\mathrm{calc}}=2655.0$\,MHz and $B_{0,\,\mathrm{exp}}=2656.393$\,MHz (0.05\,\%).
Applying the ratio $B_\mathrm{exp}/B_\mathrm{calc}$(\ce{HC3P}) to the calculated 
$B_0=2733.1$~MHz
constant of \hcsp, a best estimate value of $B_0=2734.5$~MHz is obtained. 
The uncertainties of these best estimates may well be within 1\,MHz, more than adequate for 
experimental searches by microwave techniques. Both ions are calculated to be very polar, with dipole moments of 3.26\,D for \hcop and 1.73\,D for \hcsp the CCSD(T)/aug-cc-pwCVQZ level of theory.

\section*{Acknowledgement(s)}

This work has been supported via Collaborative Research Centre 956, 
sub-project B2, funded by the Deutsche Forschungsgemeinschaft 
(DFG; project ID 184018867) and DFG SCHL 341/15-1 
(“Cologne Center for Terahertz Spectroscopy”). 
We thank the Regional Computing Center of the Universität zu Köln (RRZK) for providing computing time on the DFG-funded High Performance Computing (HPC) system CHEOPS. We gratefully acknowledge the support of Radboud University and of the Nederlandse Organisatie voor Wetenschappelijk Onderzoek (NWO), for providing the required beam time at the FELIX Laboratory and the skillful assistance of the FELIX staff. The research leading to this publication has been supported by the Project CALIPSOplus under the Grant No. 730872 from the EU Framework Programme for Research and Innovation HORIZON 2020. M.C.M and K.L.K.L. acknowledge financial support from NASA grant 80NSSC18K0396.

Lastly, we thank Marie-Aline Martin-Drumel for support during the \hcop\ campaign and a careful reading of the manuscript as well as Mr. Bryan Adams (``Summer of 69'') for many entertaining moments during the spectroscopic study of the \hcsp ion, lending inspiration to the title of this work.

%\section*{Disclosure statement}
%An unnumbered section, e.g.\ \verb"\section*{Disclosure statement}", may be used to declare any potential conflict %of interest and included \emph{in the non-anonymous version} before any Notes or References, after any %Acknowledgements and before any Funding information.

%\section*{Funding}
%An unnumbered section, e.g.\ \verb"\section*{Funding}", may be used for grant details, etc.\ if required and %included \emph{in the non-anonymous version} before any Notes or References.

\pagebreak

\bibliographystyle{tfo}
\bibliography{sthorwirth_bibdesk, LIRTRAP}

\pagebreak

\section{Appendices}
\label{appendix_a}

Internal coordinates for the Ne-complexes of \hcop (calculated at the fc-CCSD(T)/aug-cc-pVTZ level) and \hcsp 
(fc-CCSD(T)/aug-cc-pV(T+$d$)Z)  from fully relaxed structural calculations. Bond lengths are given in \AA, angles in degrees.

\subsection{Ne$-$\hcop , fc-CCSD(T)/aug-cc-pVTZ}

\subsubsection{Ne$-$\ce{HC3O+}, linear form}
\scriptsize

\begin{verbatim}
O
C 1 r1
X 2 rd 1 a90
C 2 r2 3 a90 1 d180
X 4 rd 2 a90 3 d0
C 4 r3 5 a90 2 d180
X 6 rd 4 a90 5 d0
H 6 r4 7 a90 4 d180
X 8 rd 6 a90 7 d0
NE 8 r5 9 a90 6 d180

r1   =        1.129675503855228
rd   =        1.000000409314806
a90  =       90.000000000000000
r2   =        1.345852682859310
d180 =      180.000000000000000
d0   =        0.000000000000000
r3   =        1.219943167034099
r4   =        1.076271649089833
r5   =        2.191966102105850    
\end{verbatim}

\subsubsection{\ce{HC3O+}$-$Ne, non-linear (T-) form}
\scriptsize

\begin{verbatim}
H
C 1 r1
X 2 rd 1 a90
C 2 r2 3 a1 1 d180
X 4 rd 2 a90 3 d0
C 4 r3 5 a2 2 d180
X 6 rd 4 a90 5 d0
O 6 r4 7 a3 4 d180
NE 8 r5 6 a4 7 d0

r1   =        1.075618393685392
rd   =        1.000002251233508
a90  =       90.000000000000000
r2   =        1.219547134225782
a1   =       89.987605394672087
d180 =      180.000000000000000
d0   =        0.000000000000000
r3   =        1.346336487375038
a2   =       90.223210508956072
r4   =        1.129411452271238
a3   =       90.218170030502677
r5   =        3.167121007415886
a4   =       69.294364994274972    
\end{verbatim}

\subsubsection{\ce{HC3O+}$-$Ne, linear form}
\scriptsize

\begin{verbatim}
O
C 1 r1
X 2 rd 1 a90
C 2 r2 3 a90 1 d180
X 4 rd 2 a90 3 d0
C 4 r3 5 a90 2 d180
X 6 rd 4 a90 5 d0
H 6 r4 7 a90 4 d180
X 1 rd 2 a90 3 d0
NE 1 r5 9 a90 2 d180

r1   =        1.129522777467840
rd   =        1.000000613972272
a90  =       90.000000000000000
r2   =        1.346337295083560
d180 =      180.000000000000000
d0   =        0.000000000000000
r3   =        1.219669287299509
r4   =        1.075691499375670
r5   =        2.941664123884200
\end{verbatim}

\subsection{Ne$-$\hcsp , fc-CCSD(T)/aug-cc-pV(T+d)Z}

\subsubsection{Ne$-$\ce{HC3S+}, linear form}
\scriptsize

\begin{verbatim}
S
C 1 r1
X 2 rd 1 a90
C 2 r2 3 a90 1 d180
X 4 rd 2 a90 3 d0
C 4 r3 5 a90 2 d180
X 6 rd 4 a90 5 d0
H 6 r4 7 a90 4 d180
X 8 rd 6 a90 7 d0
NE 8 r5 9 a90 6 d180

r1   =        1.508396620643273
rd   =        1.000000000000000
a90  =       90.000000000000000
r2   =        1.342701023969591
d180 =      180.000000000000000
d0   =        0.000000000000000
r3   =        1.224200791251630
r4   =        1.073475625100924
r5   =        2.241879594490575
\end{verbatim}

\subsubsection{\ce{HC3S+}$-$Ne, non-linear (T-) form}
\scriptsize

\begin{verbatim}
H
C 1 r1
X 2 rd 1 a90
C 2 r2 3 a1 1 d180
X 4 rd 2 a90 3 d0
C 4 r3 5 a2 2 d180
X 6 rd 4 a90 5 d0
S 6 r4 7 a3 4 d180
NE 8 r5 6 a4* 7 d0

r1   =        1.073113918449037
rd   =        1.000001023287332
a90  =       89.000000000000000
r2   =        1.223905552313753
a1   =       91.009405591888765
d180 =      180.000000000000000
d0   =        0.000000000000000
r3   =        1.343046564596132
a2   =       91.035318504696349
r4   =        1.508072750853718
a3   =       91.003095953954897
r5   =        3.668256915522843
a4   =       60.467779413084898    
\end{verbatim}

\subsubsection{\ce{HC3S+}$-$Ne, linear form}
\scriptsize

\begin{verbatim}
S
C 1 r1
X 2 rd 1 a90
C 2 r2 3 a90 1 d180
X 4 rd 2 a90 3 d0
C 4 r3 5 a90 2 d180
X 6 rd 4 a90 5 d0
H 6 r4 7 a90 4 d180
X 1 rd 2 a90 3 d0
NE 1 r5 9 a90 2 d180

r1   =        1.508253279164069
rd   =        1.000000613972272
a90  =       90.000000000000000
r2   =        1.343211324073055
d180 =      180.000000000000000
d0   =        0.000000000000000
r3   =        1.223930047649706
r4   =        1.073146611305811
r5   =        3.177832307922192    
\end{verbatim}

\end{document}